\documentclass[conference]{IEEEtran}
\IEEEoverridecommandlockouts
\usepackage{amsmath,amssymb,amsfonts}
\usepackage{algorithmic}
\usepackage{graphicx}
\usepackage{multirow}
\usepackage{booktabs}
\usepackage{enumitem}
\usepackage[small]{caption}
\usepackage{textcomp}
\usepackage{xcolor,soul}
\usepackage{caption}
\usepackage[square, comma, sort&compress,numbers]{natbib}
\def\BibTeX{{\rm B\kern-.05em{\sc i\kern-.025em b}\kern-.08em
    T\kern-.1667em\lower.7ex\hbox{E}\kern-.125emX}}

\graphicspath{{Figures/}{Figures/BD/}{Figures/BVI-AVP/}}

\usepackage{tikz}
\usepackage{subcaption}
\usetikzlibrary{matrix,chains,positioning,decorations.pathreplacing,arrows,spy}
\tikzset{every picture/.style={font issue=\footnotesize},
         font issue/.style={execute at begin picture={#1\selectfont}}
        }
\usetikzlibrary{shapes.geometric}

\newlength{\Oldarrayrulewidth}

\begin{document}

\title{A Subjective Study on Videos at Various Bit Depths}
\author{Alex Mackin, Di Ma, Fan Zhang and David Bull \\
\textit{Bristol Vision Institute, University of Bristol, Bristol, UK, BS1 5DD.}\\
\{A.Mackin, Di.Ma, Fan.Zhang, Dave.Bull\}@bristol.ac.uk}

\maketitle

\begin{abstract}
Bit depth adaptation, where the bit depth of a video sequence is reduced before transmission and up-sampled during display, can potentially reduce data rates with limited impact on perceptual quality. In this context, we conducted a subjective study on a UHD video database, BVI-BD, to explore the relationship between bit depth and visual quality. In this work, three bit depth adaptation methods  are investigated, including linear scaling, error diffusion, and a novel adaptive Gaussian filtering approach. The results from a subjective experiment indicate that above a critical bit depth, bit depth adaptation has no significant impact on perceptual quality, while reducing the amount information that is required to be transmitted. Below the critical bit depth, advanced adaptation methods can be used to retain `good' visual quality (on average) down to around 2 bits per color channel for the outlined experimental setup - a large reduction compared to the typically used 8 bits per color channel. A selection of image quality metrics were subsequently bench-marked on the subjective data, and analysis indicates that a bespoke quality metric is required for bit depth adaptation.
\end{abstract}

%\vspace{-1em}
%\vspace{-0.115cm}

\section{Introduction}
\label{sec:intro}

Recent video formats~\cite{2020} expand the video parameter space to deliver more immersive experiences~\cite{Bull2021}
% \textcolor{red}{XXX Can we cite the new book please: D. Bull and F. Zhang, Intelligent Image and Video Compression: Communicating pictures 2e,  Elsevier, 2021. XXXX}[fixed]
through increased spatial resolution, frame rate and/or dynamic range/color gamut (up to 12 bits per color channel~\cite{2020}). The increased data bandwidth associated with these formats has been addressed in the latest video coding standards, including MPEG Versatile Video Coding (VVC) \cite{s:VVC} and AOMedia AV1 \cite{w:AV1}.

Further coding gains can be achieved (even over the latest standards) by exploiting the properties of the human visual system (HVS) through spatial and temporal resolution adaptation~\cite{afonso2018video}. More recently, similar methods have been proposed by extending the adaptation to the bit depth. It is noted that in digital video applications, a bit depth of 8 ($2^8$ = $256$ pixel values) has been typically used such that each pixel is represented as a byte of information. While this representation is convenient for data processing, it is not necessarily the most efficient in terms of visual perception. Bit depth adaptation refers to the case where  content is captured at a perceptually lossless bit depth (e.g. 16 bits) and then an optimal bit depth is determined during the video coding process. The full bit depth can be then recovered via up-sampling after decoding for display. This approach has been reported to provide consistent coding gains over standard video coding algorithms \cite{zhang2019vistra2,ma2020gan}.

Previously we demonstrated a content dependent relationship between visual quality and both spatial~\cite{Mackin2018(Submitted)} and temporal~\cite{Mackin2015,mackin2018study,mackin2020} resolutions. This indicates, that given a suitable quality metric~\cite{zhang2017frame,mackin2018srqm,madhusudana2020capturing}, resolution could be selected in the context of compression through quantisation resolution optimisation, which trades-off compression artifacts and the distortion introduced through resolution adaption \cite{ma2019perceptually,ma2020cvegan}. However, the relationship between bit depth and perceptual quality has not yet been properly investigated.

The sensitivity of the HVS to changes in intensity (luminance) can be explained using Just-Noticeable Differences (JND)~\cite{barten1999contrast} - the intensity delta ($\Delta I$) required for a stimulus to be perceptible against a uniform background. This delta increases with luminance under a power-law relationship. In video applications, this non-linear effect is corrected using gamma correction~\cite{Poynton2012}, such that $\Delta I$ is roughly constant between all valid pixel values. The contrast sensitivity function extends the JND principle to structured stimuli, and demonstrates that $\Delta I$ is a function of spatial and temporal frequencies~\cite{barten1999contrast} - demonstrating content dependence. False edges (banding) become perceptible if the luminance step between pixel values is greater than $\Delta I$. Therefore the higher the contrast ratio of the display (difference between brightest and darkest elements), the greater the number of pixel values required. This explains why HDR (High Dynamic Range) displays, with contrast ratios in excess of 20,000:1, have bit depths of 10/12bits~\cite{2020}, rather than the traditional 8 bits~\cite{709}. 

In this paper, we extend our previous work to explore the impact of bit depth variations on perceptual quality. While this investigation is driven by the recent emergence of HDR and high bit-depth formats~\cite{narwaria2015high}, it is equally applicable to traditional formats where bit depth adaptation has been reported to provide significant improvement for video compression \cite{zhang2019vistra2}. To conduct this study, a UHD video database was developed, BVI-BD, which contains 12 source sequences with a native bit depth of 10. Each sequence was subsequently down-sampled to four lower test bit depths and then up-sampled to the original bit depth based on three different adaptation strategies. The subjective results show that there is a critical bit depth (around 6 bits) for our experimental setup, below which bit depth adaptation starts to introduce evident visual artifacts. Based on these subjective evaluations, we also evaluated the performance of a selection of popular full reference image quality metrics in the context of bit depth variations.

The remainder of this paper is organized as follows. Section~\ref{sec:database} describes the source and test sequences used in the subjective experiment and outlines the bit depth adaption methods to be explored. Section~\ref{sec:ESM} presents the experimental methodology. The subjective results are then analysed in Section~\ref{sec:res} and popular quality metrics benchmarked. Section \ref{sec:conclusion} concludes the paper and outlines steps for future work.

\begin{table*}[htbp!]
\footnotesize
\setlength{\tabcolsep}{0pt}
\renewcommand{\arraystretch}{0}
  \centering  
  \begin{tabular}{cccccc}

\includegraphics[width=0.16\textwidth]{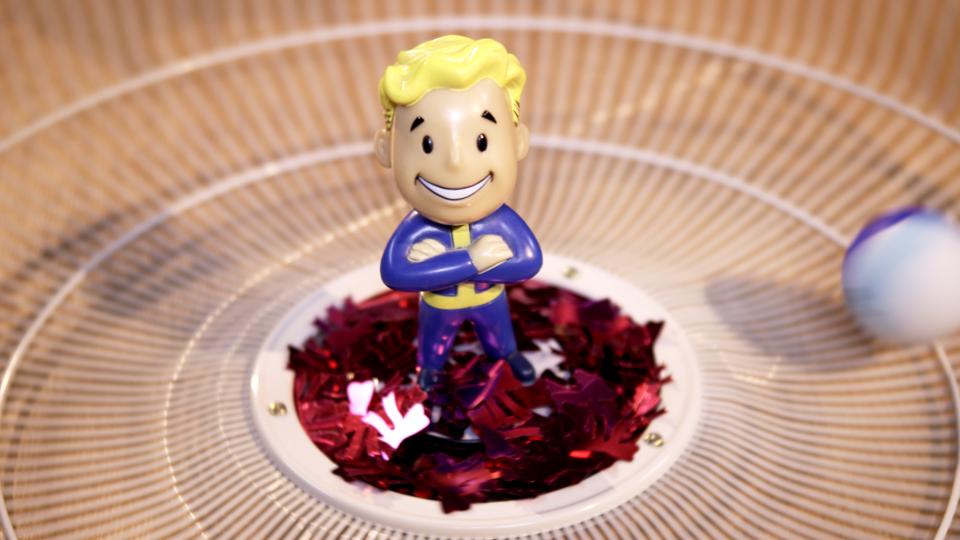}&\includegraphics[width=0.16\textwidth]{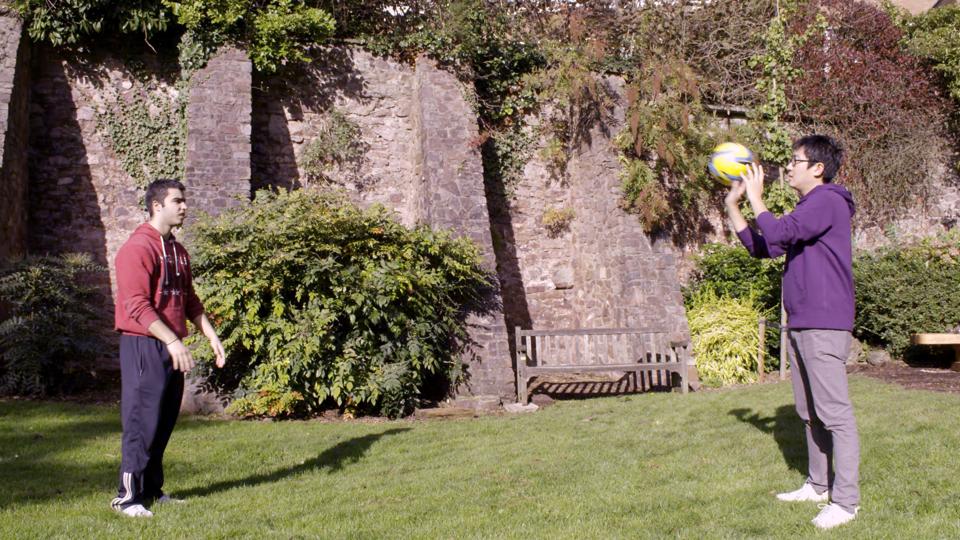}&\includegraphics[width=0.16\textwidth]{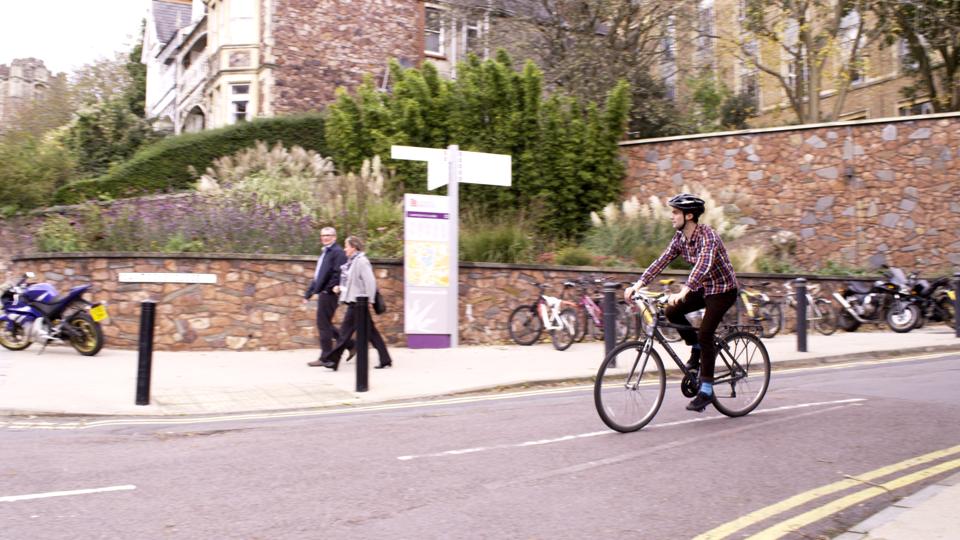}&\includegraphics[width=0.16\textwidth]{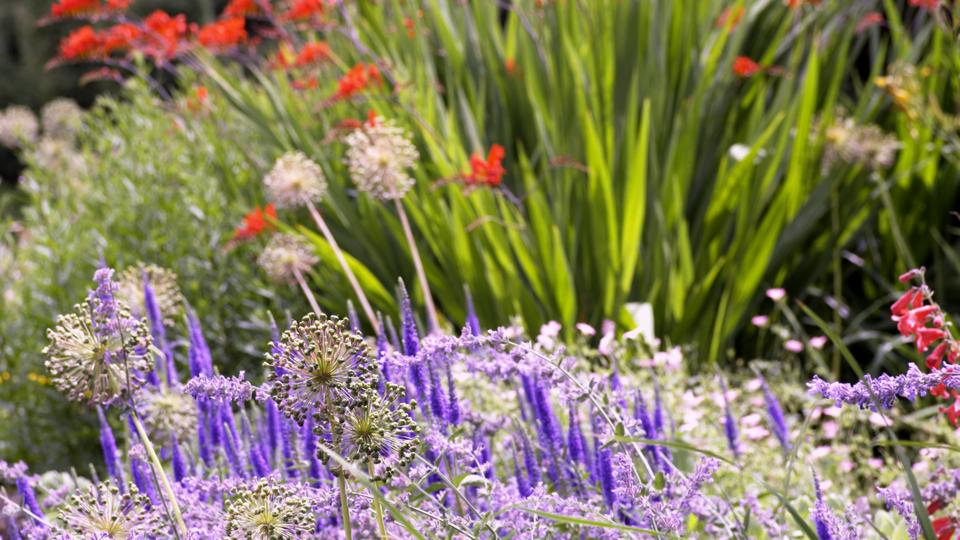}&\includegraphics[width=0.16\textwidth]{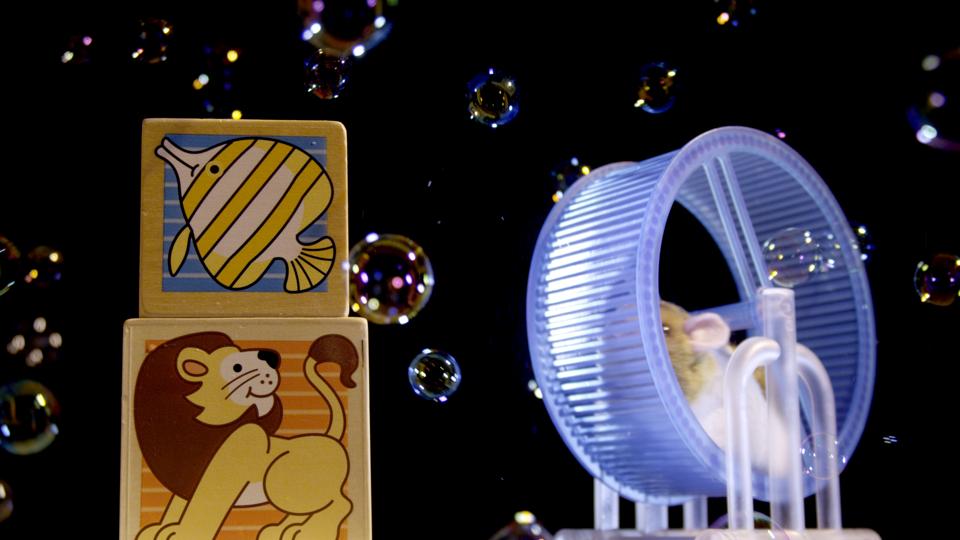}&\includegraphics[width=0.16\textwidth]{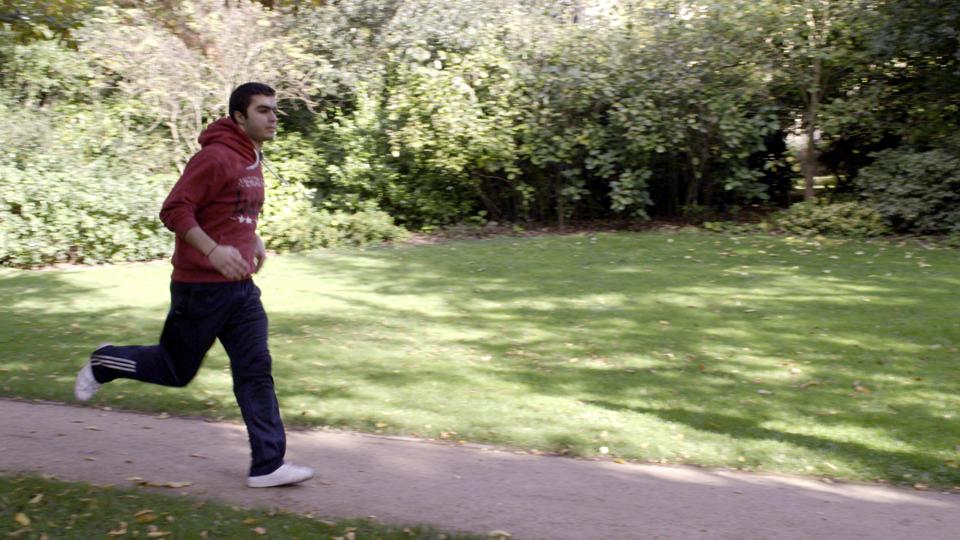}\\
Bobblehead&Catch&Cyclist&Flowers&Hamster&Joggers\\

\includegraphics[width=0.16\textwidth]{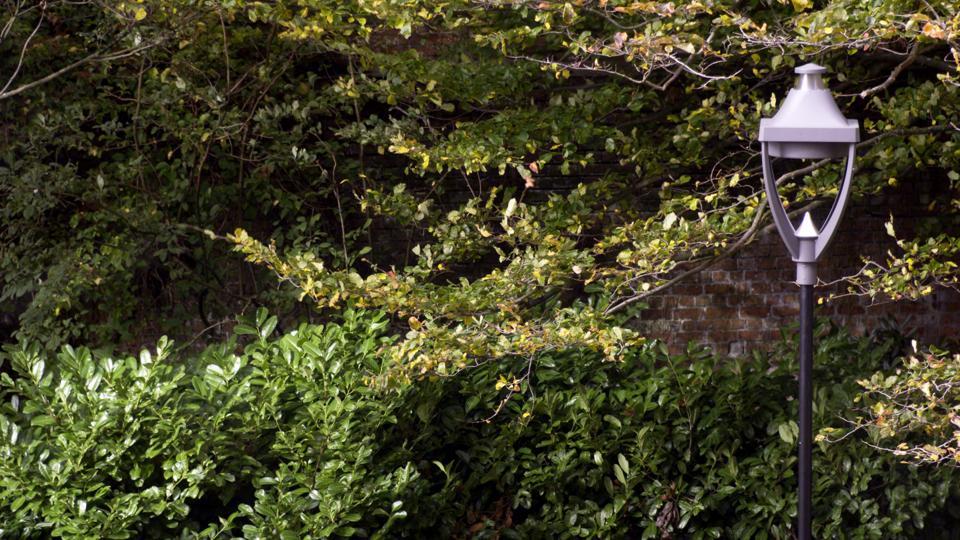}&\includegraphics[width=0.16\textwidth]{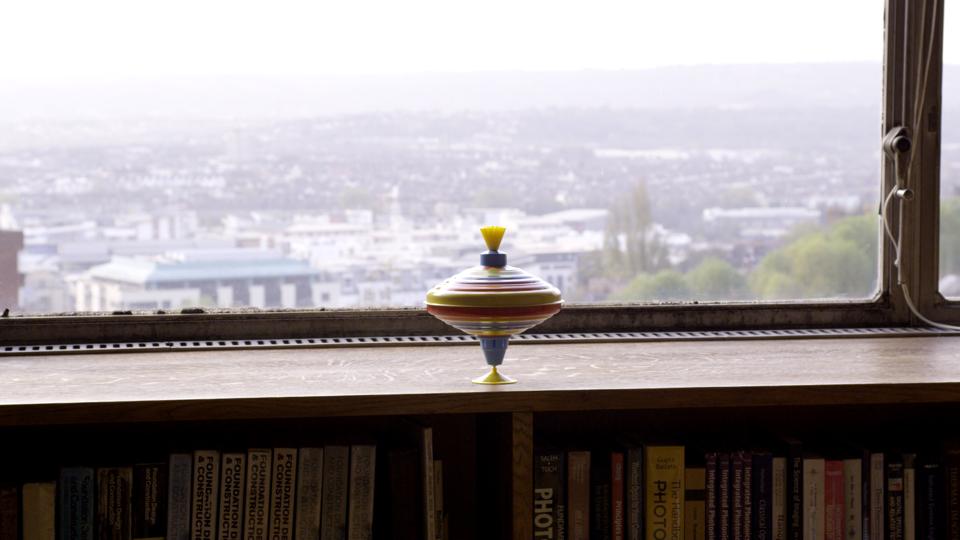}&\includegraphics[width=0.16\textwidth]{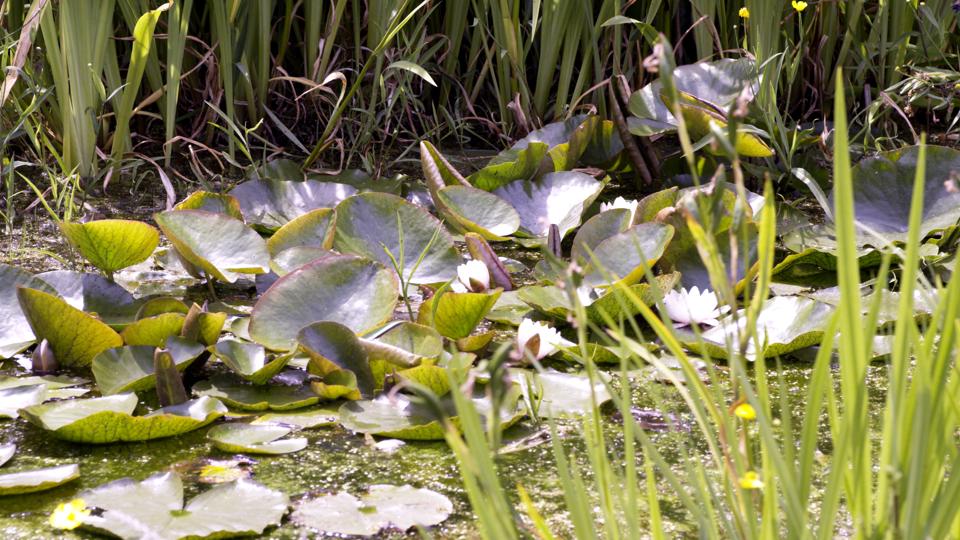}&\includegraphics[width=0.16\textwidth]{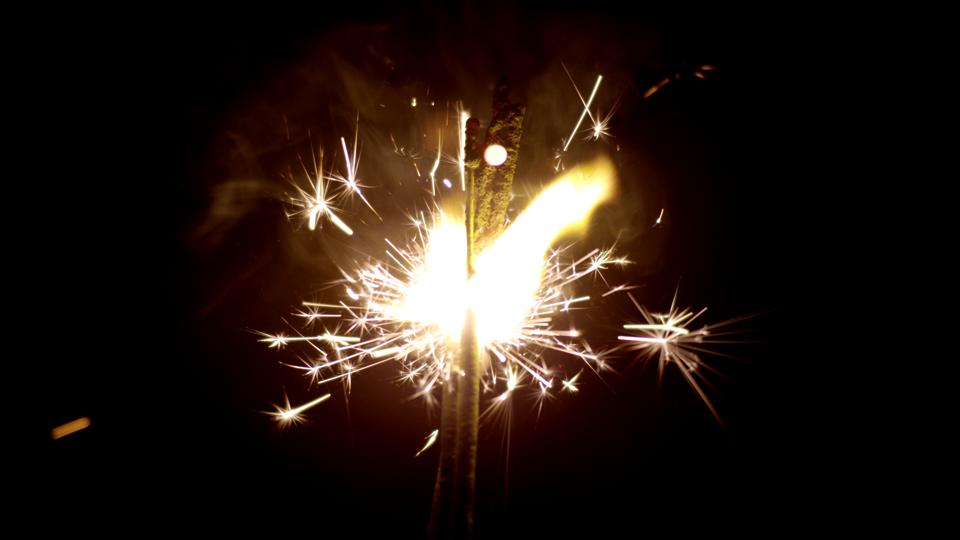}&\includegraphics[width=0.16\textwidth]{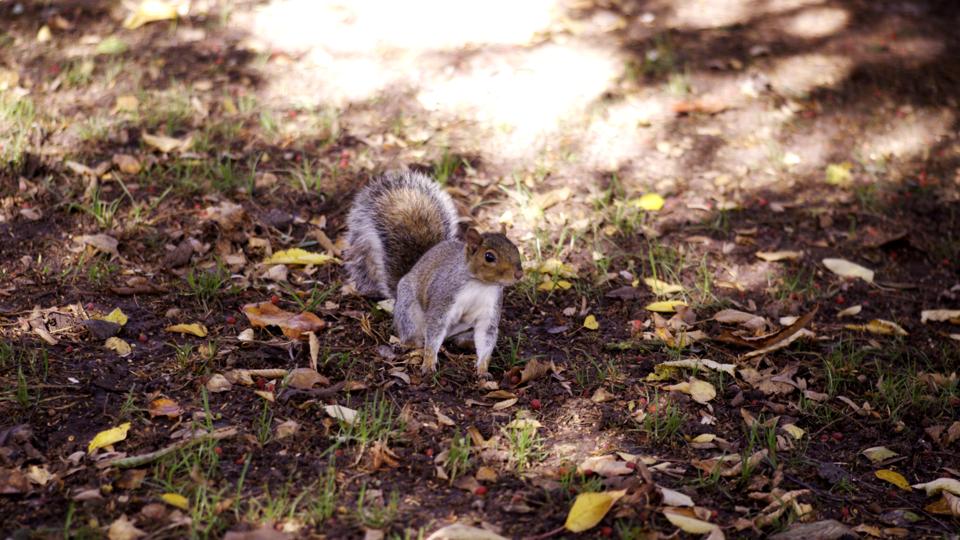}&\includegraphics[width=0.16\textwidth]{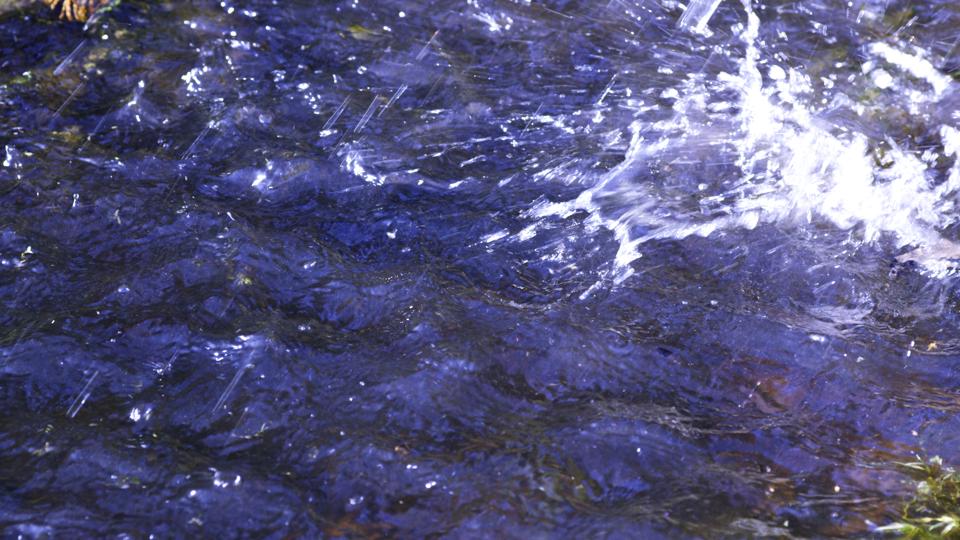}\\
Lamppost&Library&Pond&Sparkler&Squirrel&Water\\

\end{tabular}
\captionof{figure}{A sample frame from each of 12 source sequences in the BVI-BD video database.}
  \label{fig:sequences}
\end{table*}

\section{BVI-BD Database}
\label{sec:database}
This section provides an overview of the video database, BVI-BD, and the test sequences used in the experiments.

\subsection{Source Sequences}

The Bristol Vision Institute Bit Depth (BVI-BD) video database\footnote{All the test sequences and subjective results are available to download from https://fan-aaron-zhang.github.io/BVI-BD/} contains 12 unique sequences (with example frames from BVI-BD shown in Fig.~\ref{fig:sequences}). Each sequence was captured using a RED Epic-X video camera at a 3840$\times$2160p spatial resolution and a frame rate of 120 fps (360$^{\circ}$ shutter angle). The sequences were down-sampled by frame averaging~\cite{mackin2019frame} to 60fps and graded in BT.2020 color space~\cite{2020} using REDCINE-X at 10 bits per color channel (bpc). The experiment encompasses a range of content types including static (e.g. Bobblehead, Hamster) and dynamic (e.g. Sparkler, Water) textures. Each sequence was cut to 5s in duration, as recommended in~\cite{Moss2016} for single stimulus video quality assessment.

\subsection{Test Sequences}

The 12 source sequences were down-sampled to bit depths of 8, 6, 4 and 2 bits, and subsequently up-sampled to the original bit depth of 10 bits for display. The original 10 bit video was included as a reference. This process emulates bit depth adaptation. Three separate methods for bit depth adaptation were investigated: (i) linear down-sample with linear up-sample (Linear), (ii) error diffusion down-sample with linear up-sample (Error Diffusion), and (iii) error diffusion down-sample with an adaptive Gaussian filter applied after linear up-sampling (the $\ell_2$-optimized $\sigma$ is calculated per sequence for the purpose of this paper). This results in 156 (12$\times$4$\times$3+12) test sequences (including the original 10 bit sources). The three tested adaptation methods are outlined in further detail below.

\begin{table}[htbp!]
\footnotesize
\setlength{\tabcolsep}{-3pt}
\renewcommand{\arraystretch}{0}
  \centering  
  \begin{tabular}{cc}
\begin{tikzpicture}[spy using outlines={circle,magnification=3,size=2cm, every spy on node/.append style={thick},connect spies}]
	\node[anchor=south west] (img) at (0,0)
    {\includegraphics[clip,width=0.5\columnwidth]{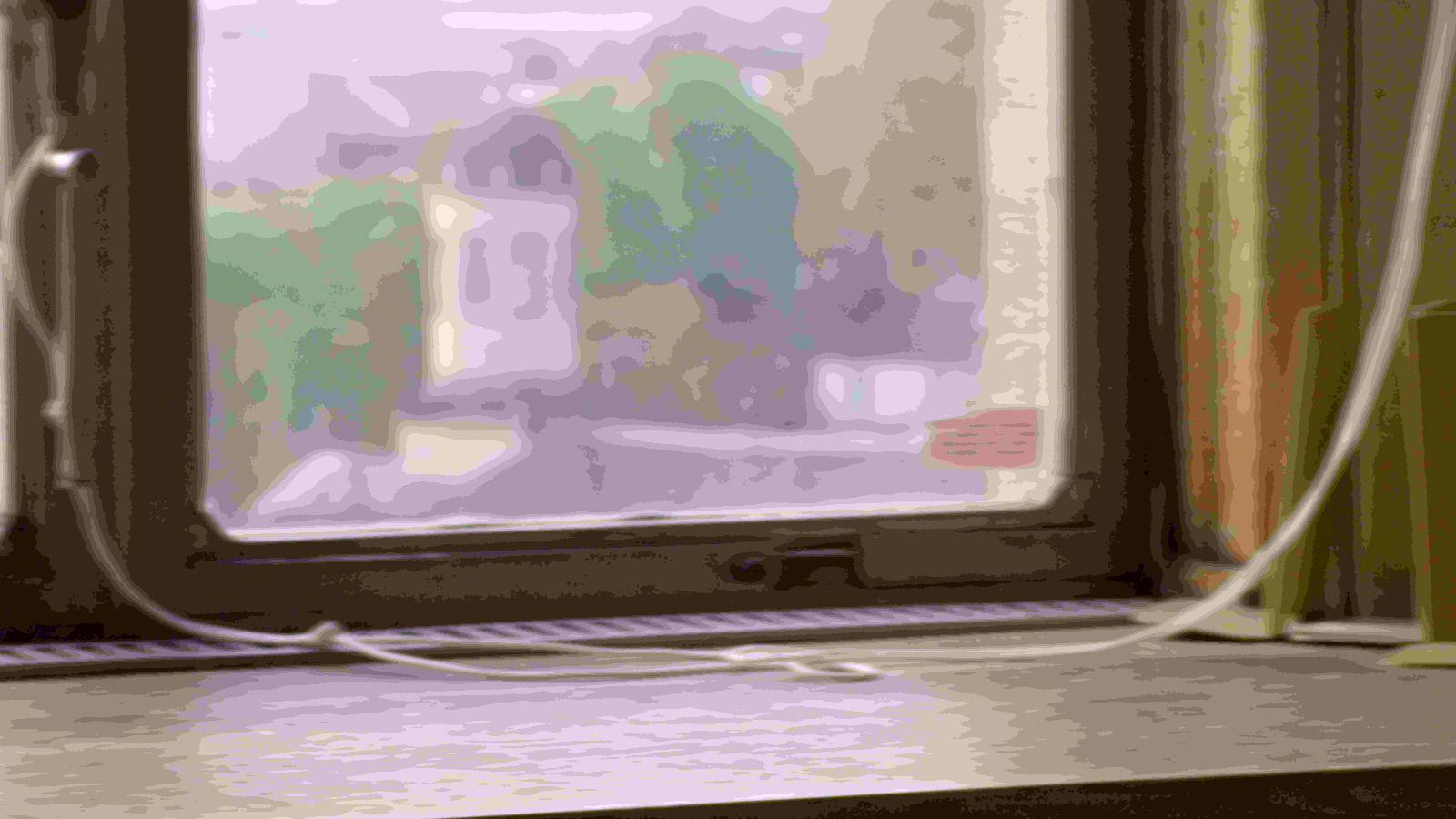}};
\spy on (1.45,1.9) in node [left,thick] at (4.35,1.35);
\end{tikzpicture}
&
\begin{tikzpicture}[spy using outlines={circle,magnification=3,size=2cm, every spy on node/.append style={thick},connect spies}]
	\node[anchor=south west] (img) at (0,0)
    {\includegraphics[clip,width=0.5\columnwidth]{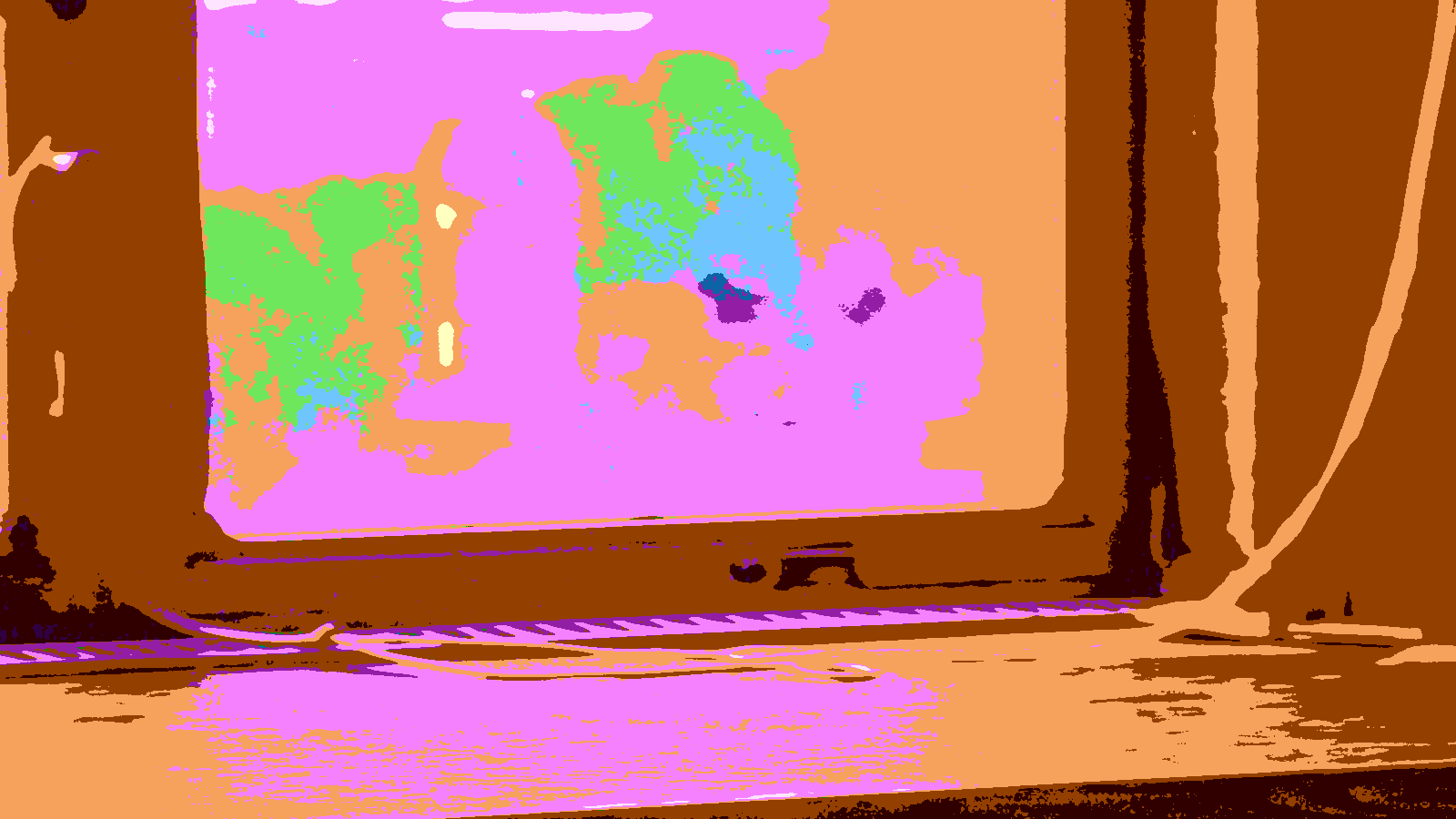}};
\spy on (1.45,1.9) in node [left,thick] at (4.35,1.35);
\end{tikzpicture}\\
(a)&(b)\\
\end{tabular}
\captionof{figure}{A frame from Library (10 bit) converted to (a) 4 bit (PSNR: 28.4dB) and (b) 2 bit (PSNR: 13.0dB) using linear scaling.}
\label{fig:lbd}
\end{table}

\textbf{Linear: } Linear scaling is the simplest method to convert between bit depths $a$ and $b$, and is defined as follows:
\begin{equation}
\label{eq:linear}
I_b(x,y,c) =  \left\lfloor (M_a/M_b)\cdot I_a(x,y,c)\right\rceil
\end{equation}
where $M_n$ is the maximum value for bit depth $n$ (for unsigned integer this would be $2^{n-1}$). $I(x,y,c)$ is the pixel value (intensity) at location $x$, $y$ for color channel $c$. $\left\lfloor x \right\rceil$ indicates the rounding operation (nearest integer). A major drawback of this method is the introduction of contouring artifacts and color misrepresentations at low bit depths as shown in Fig.~\ref{fig:lbd}.

\textbf{Error Diffusion:} Error diffusion (or image dithering)~\cite{steinberg1976adaptive} attempts to overcome the limitations of linear scaling by propagating errors across multiple neighbouring pixels, and therefore decorrelating the quantisation noise from the input image signal. This process leads to a visual perception similar to additive random noise, instead of the structured distortions and color misrepresentations associated with linear scaling. 

The first stage of error diffusion is to calculate the normalized error under a linear scaling (Eq.~\ref{eq:linear}), which is:
\begin{equation}
e(x,y,c) = \left(I_a(x,y,c) - I_b(x,y,c)\right)/M_a
\end{equation}

The current pixel value is converted using the linear scaling method in Eq.~\ref{eq:linear}, while the neighbouring pixels are modulated by a factor ($K$) of the normalized error as follows:
\begin{equation}
I_b(x+i,y+j,c) = I_a(x+i,y+j,c)  + K \cdot e(x,y,c)\\
\end{equation}
Starting from the top-left corner, a horizontal scanning pattern is typically employed until every pixel has been updated. This operation is applied to each color channel independently.

\begin{figure}[htbp!] 
\centering
\begin{tikzpicture}[scale=1.4]
\fill [orange!50] (0,2) rectangle (1,3);
\draw[dashed] (-2,2) rectangle (-1,3);
\draw[dashed] (-1,2) rectangle (0,3);
\draw (0,2) rectangle (1,3) node[pos=.5] {};
\draw (1,2) rectangle (2,3) node[pos=.5] {\Large$\frac{5}{32}$};
\draw (2,2) rectangle (3,3) node[pos=.5] {\Large$\frac{3}{32}$};

\draw (-2,1) rectangle (-1,2) node[pos=.5] {\Large$\frac{1}{16}$};
\draw (-1,1) rectangle (0,2) node[pos=.5] {\Large$\frac{1}{8}$};
\draw (0,1) rectangle (1,2) node[pos=.5] {\Large$\frac{5}{32}$};
\draw (1,1) rectangle (2,2) node[pos=.5] {\Large$\frac{1}{8}$};
\draw (2,1) rectangle (3,2) node[pos=.5] {\Large$\frac{1}{16}$};

\draw[dashed] (-2,0) rectangle (-1,1);
\draw (-1,0) rectangle (0,1) node[pos=.5] {\Large$\frac{1}{16}$};
\draw (0,0) rectangle (1,1) node[pos=.5] {\Large$\frac{3}{32}$};
\draw (1,0) rectangle (2,1) node[pos=.5] {\Large$\frac{1}{16}$};
\draw[dashed] (2,0) rectangle (3,1);
\end{tikzpicture}
\caption{The Sierra coefficient matrix (current pixel is shaded orange).}
\label{fig:Sierra}
\end{figure}

\begin{table}[htbp!]
\footnotesize
\setlength{\tabcolsep}{-3pt}
\renewcommand{\arraystretch}{0}
  \centering  
  \begin{tabular}{cc}
\begin{tikzpicture}[spy using outlines={circle,magnification=3,size=2cm, every spy on node/.append style={thick},connect spies}]
	\node[anchor=south west] (img) at (0,0)
    {\includegraphics[clip,width=0.5\columnwidth]{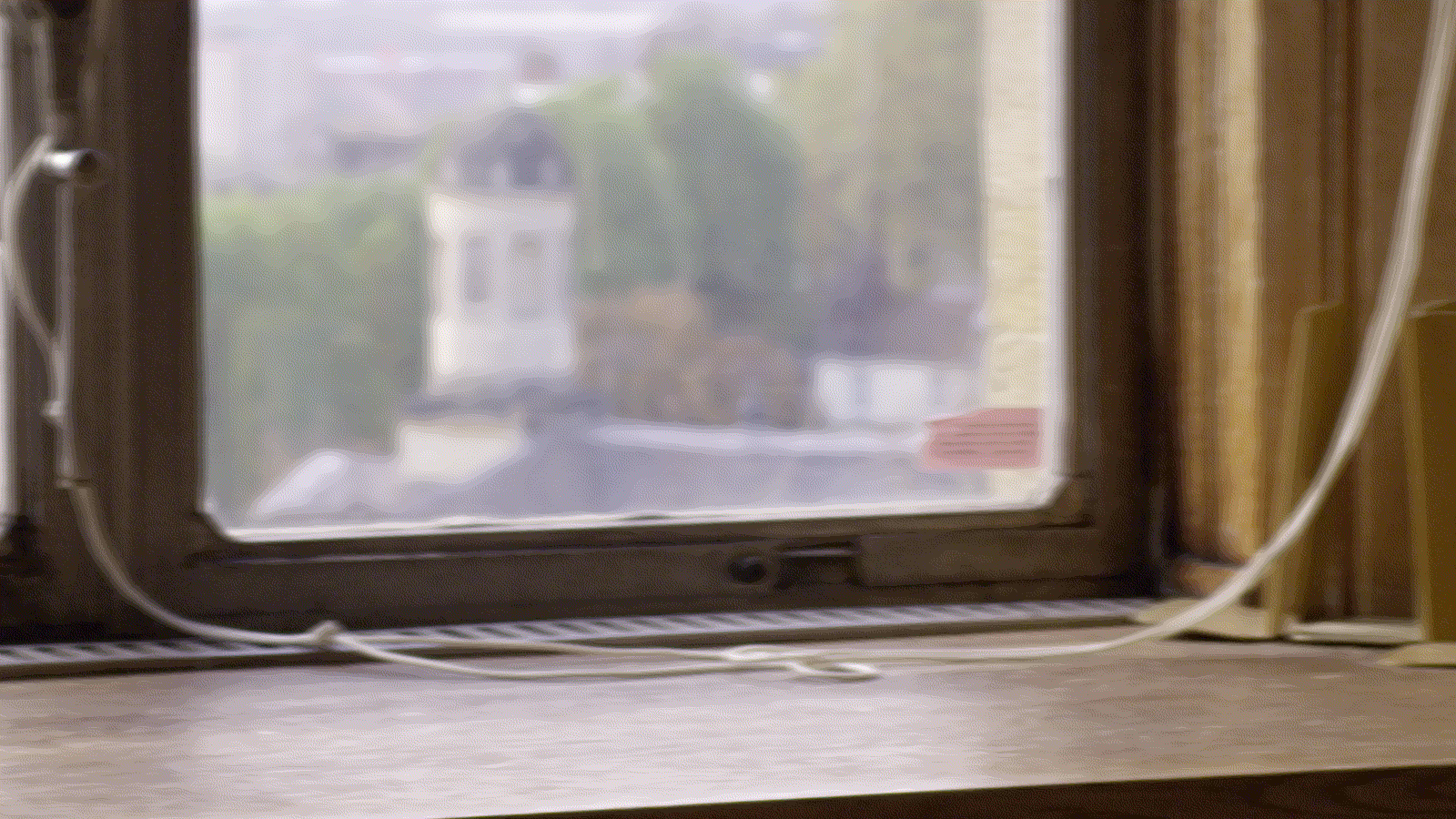}};
\spy on (1.45,1.9) in node [left,thick] at (4.35,1.35);
\end{tikzpicture}
&
\begin{tikzpicture}[spy using outlines={circle,magnification=3,size=2cm, every spy on node/.append style={thick},connect spies}]
	\node[anchor=south west] (img) at (0,0)
    {\includegraphics[clip,width=0.5\columnwidth]{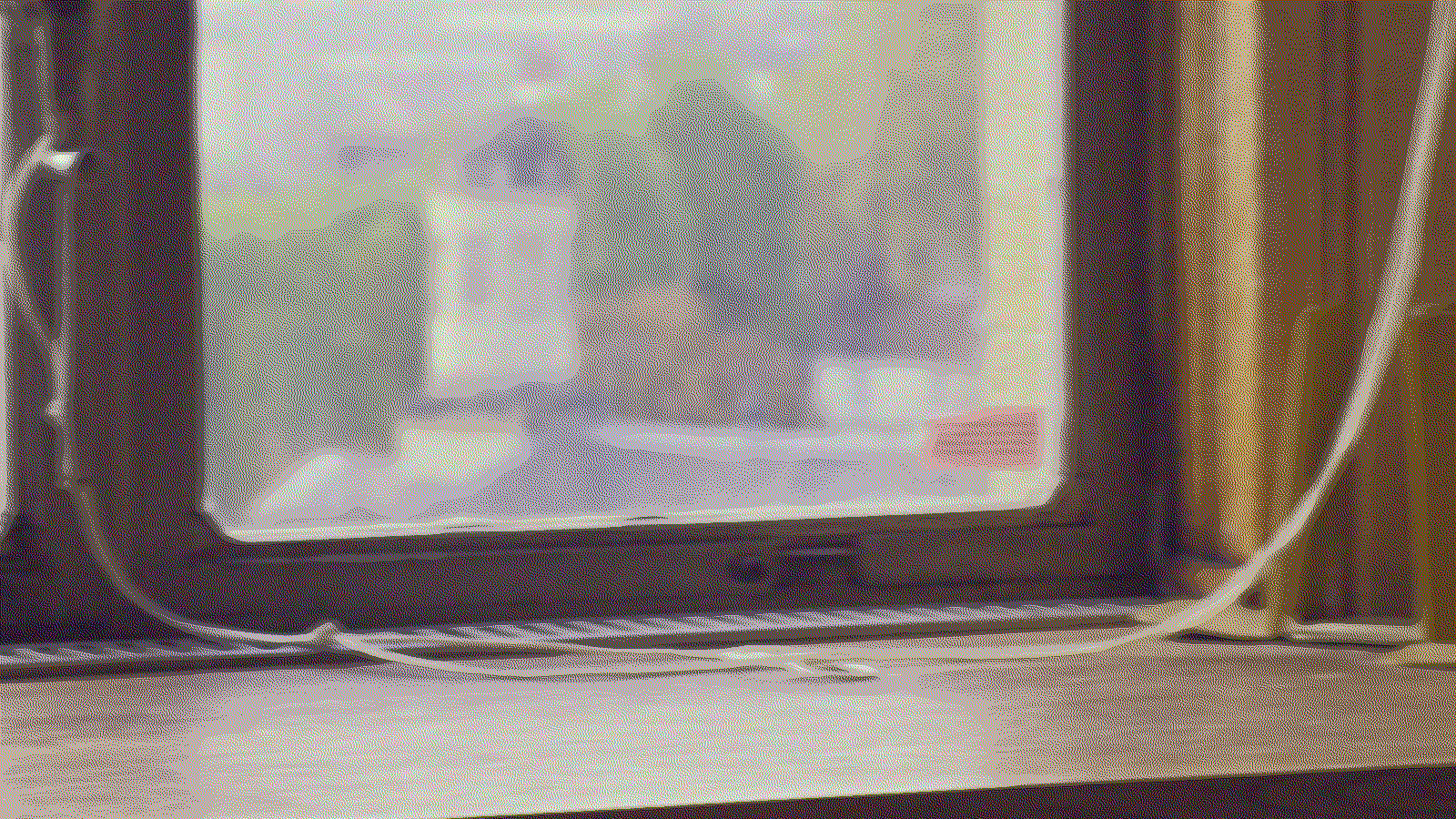}};
\spy on (1.45,1.9) in node [left,thick] at (4.35,1.35);
\end{tikzpicture}\\
(a)&(b)\\
\end{tabular}
\captionof{figure}{A frame from Library (10 bit) converted to (a) 4 bit (PSNR: 25.8dB) and (b) 2 bit (PSNR: 11.5dB) using the Sierra method described in Fig.~\ref{fig:Sierra} (applied to each color channel independently).}
\label{fig:sbd}
\end{table}

The coefficient matrix defines the modulation factor ($K$) and diffusion pattern. While Floyd-Steinberg~\cite{steinberg1976adaptive} dithering is the most popular, we found during informal study that the Sierra (Fig.~\ref{fig:Sierra}), a variant on Jarvis dithering~\cite{jarvis1976survey}, provided the highest visual quality. Sierra-Lite is a low-complexity version.

Fig.~\ref{fig:sbd} shows an example of Sierra error diffusion. While this method has a lower PSNR than linear scaling, visual quality is much higher due to retention of structure and color.

In order to reduce the number of methods investigated, we have intentionally ignored advanced scanning patterns~\cite{riemersma1998balanced}, block-based dithering~\cite{damera2001fm} and CNN-based methods~\cite{zhang2019enhanced}.

\textbf{Adaptive Gaussian Filter:} The properties of error diffusion can be exploited to more accurately recover the original image when up-sampling. Many approaches have been proposed for bit depth up-sampling~\cite{Wong1995,Xiong1999,Kite2000,Kite2000a,Neelamani2002,Li2011,Yang2015,zhang2019enhanced}, but these typically are designed for single bit imagery, do not support multiple channels (color) and/or have very high computational complexity.

We pose that the bit depth reconstruction problem can be approximated as a linear inverse problem in the form:
\begin{equation}
I_a \approx A * \left\lfloor \frac{M_b}{M_a} \cdot I_b\right\rceil
\end{equation}
where $A$ is a linear convolutional filter of any size.

The optimal solution for $A$ in terms of mean squared error is a FIR Wiener filter. However this would require the local statistics of the image to be transmitted to the decoder - increasing bit rate. Therefore we instead propose to approximate $A$ as a Gaussian kernel with bandwidth $\sigma$:
\begin{equation}
G_\sigma(i,j) = \exp\left(\frac{-(i^2+j^2)}{2\sigma^2}\right)
\end{equation}
This filter retains shape through symmetry, can be incorporated into rate-distortion optimisation processes and utilizes one only parameter per frame or per sequence ($\sigma$). Fig.~\ref{fig:agbd} shows an example of this Gaussian filter (with $\ell_2$-optimized $\sigma$). The filter is normalized to sum to unity i.e. $G_\sigma(i,j) / \sum G_\sigma$.

We hypothesize that a Gaussian kernel is a good approximation to the Wiener filter because error diffusion propagates noise pseudo-symmetrically with decaying magnitude within a small neighbourhood. During implementation, $\sigma$ can either be fixed or calculated using rate-distortion optimisation (from a candidate list). While this simple adaptive filter could be improved through local dependence and/or advanced filter kernels, for the purpose of this paper we just wanted to investigate the impact of simple post-filtering on visual quality.

\begin{table}[htbp!]
\footnotesize
\setlength{\tabcolsep}{-3pt}
\renewcommand{\arraystretch}{0}
  \centering  
  \begin{tabular}{cc}
\begin{tikzpicture}[spy using outlines={circle,magnification=3,size=2cm, every spy on node/.append style={thick},connect spies}]
	\node[anchor=south west] (img) at (0,0)
    {\includegraphics[clip,width=0.5\columnwidth]{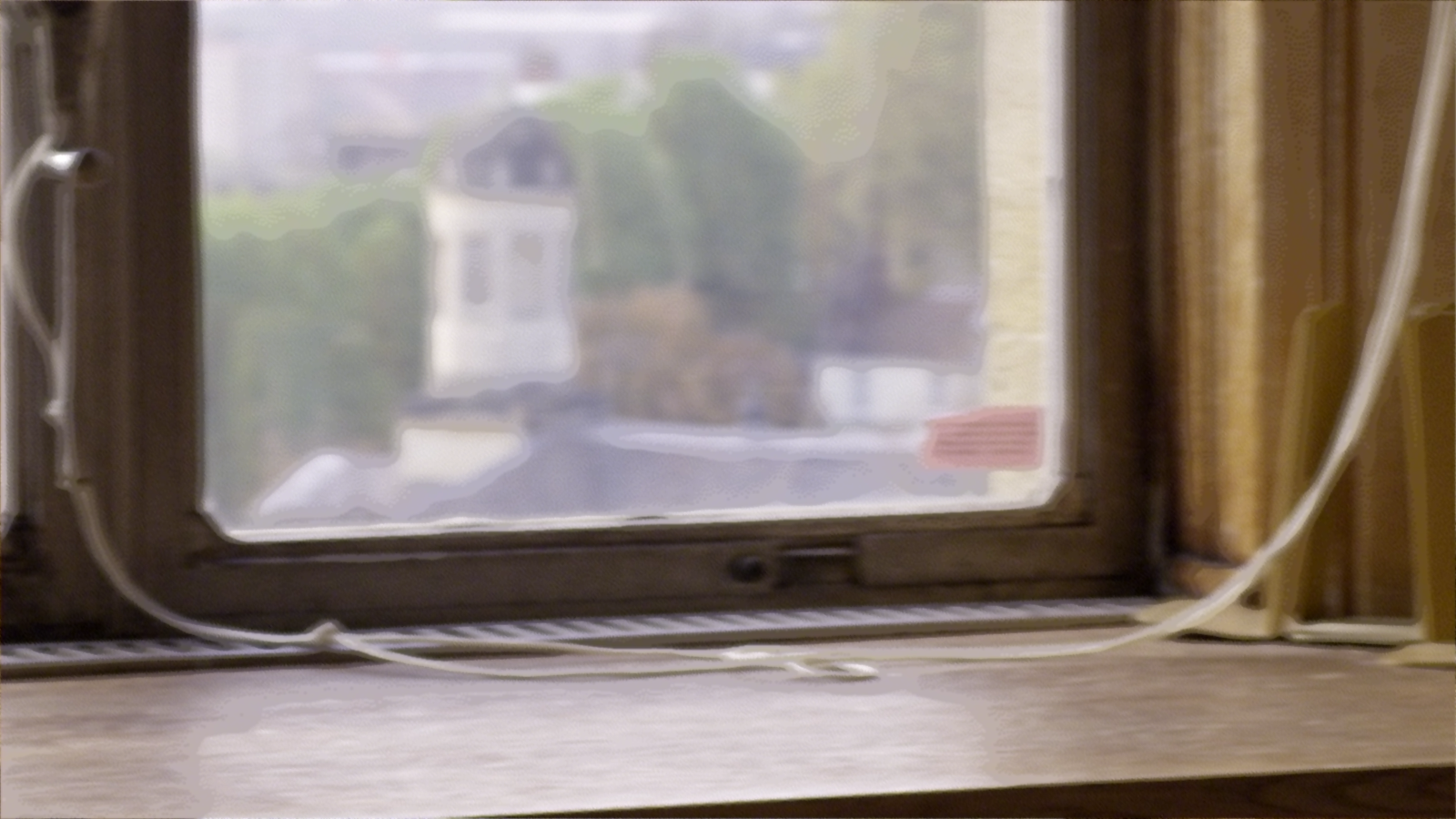}};
\spy on (1.45,1.9) in node [left,thick] at (4.35,1.35);
\end{tikzpicture}
&
\begin{tikzpicture}[spy using outlines={circle,magnification=3,size=2cm, every spy on node/.append style={thick},connect spies}]
	\node[anchor=south west] (img) at (0,0)
    {\includegraphics[clip,width=0.5\columnwidth]{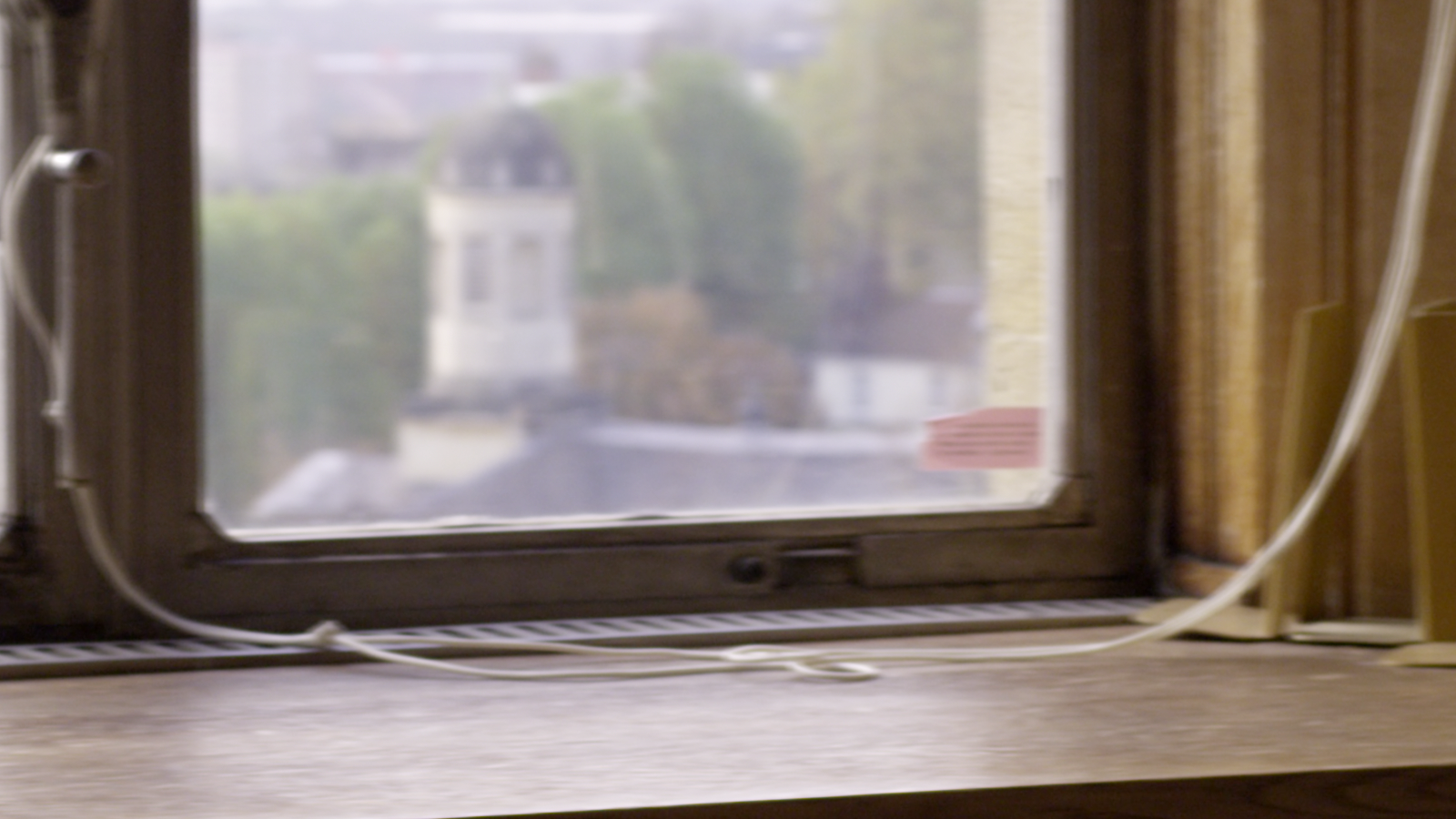}};
\spy on (1.45,1.9) in node [left,thick] at (4.35,1.35);
\end{tikzpicture}\\
(a)&(b)\\
\end{tabular}
\captionof{figure}{(a) A frame from Library (10 bit) converted to 2 bit (Sierra) and up-sampled using the adaptive Gaussian method (kernel size: $N = 2 \lceil 2 \sigma \rceil + 1$), compared to (b) the original 10 bit (PSNR: 35.1dB).}
\label{fig:agbd}
\end{table}

\section{Experimental Setup and Methodology}
\label{sec:ESM}

This section describes the experimental setup used to investigate the relationship between bit depth and visual quality.

\subsection{Experimental Setup}

A Panasonic BT-4LH310 LCD reference monitor with a peak luminance of 210cd/m$^2$ (measured using a Konica Minolta
CS-2000 spectroradiometer), a contrast ratio of 400:1, 3840$\times$2160 spatial resolution (measuring 65.4$\times$36.8cm), BT.2020 color space [1] (full range), and a refresh rate of 60 fps was used. The viewing distance was set to be 1.5H~\cite{ITUTRP2008}, while the viewing environment conformed to the home environment conditions outlined in BT.500-13~\cite{500}.

\subsection{Testing Methodology}

Each participant took part in a brief training session to acclimatize themselves with the testing process prior to the experiment. Each session lasted no longer than 30 minutes, and involved viewing the 156 test sequences. Each test involved the participant viewing a 3 second mid-level grey screen before viewing a randomly selected sequence. Participants’ then recorded their opinion on a continuous quality scale from 0 to 5~\cite{500}. A single-stimulus methodology was chosen to emulate typical viewing environments (no reference sequence).

\subsection{Participants}

Fourteen participants (8 male, 6 female) from the University of Bristol were paid to take part in each phase (both expert and non-expert viewers). The average age of participants was 27.9 $\pm$ 6.5 years. All participants had normal or corrected-to-normal color vision (verified with a Snellen chart).

\section{Results and Discussion}
\label{sec:res}

\begin{table}[htbp!]
\footnotesize
\setlength{\tabcolsep}{-2pt}
\renewcommand\arraystretch{0}
  \centering  
\vspace{-1.5ex}
  \begin{tabular}{c}
	\includegraphics[width=0.9\columnwidth]{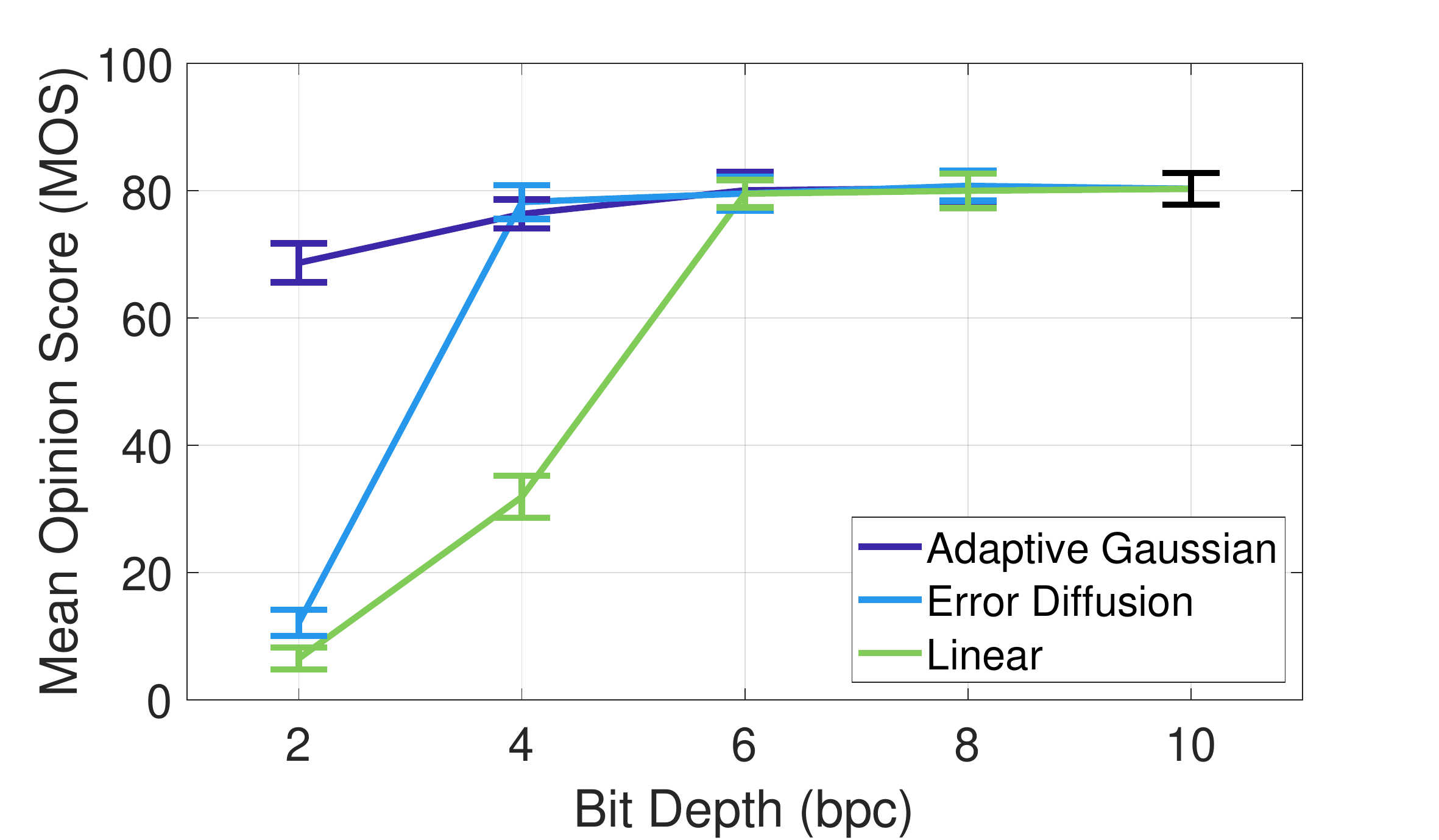}\\
\end{tabular}
\vspace{-1.5ex}
\captionof{figure}{Results from the subjective experiment showing the relationship between visual quality (MOS) and bit depth for the tested methods. Error bars show standard error of the mean ($\pm\sigma/\sqrt{n}$).}
\label{fig:MOS}
\end{table}

Mean Opinion Scores (MOS) were calculated for each test condition and linearly scaled to the range 0-100 (bad to excellent). The results from the experiment can be viewed in Fig.~\ref{fig:MOS}, in which average $\pm$ standard error MOS is reported. 

\subsection{Overall Performance}
\label{subsec:overallperformance}

The results show no significant difference (overlapping error bars) in visual quality when down-sampling to 6 bpc compared to the 10 bpc reference. The linear method exhibits a large decrease in quality below 6 bpc - predicted to be due to the color misrepresentations seen in Fig.~\ref{fig:lbd}. Error diffusion retains color consistency, and even at 4 bpc, no significant difference in visual quality can be observed  - although mean score is starting to trend downwards. The artifacts associated with error diffusion start to have a significant impact on quality at 2 bpc. The adaptive Gaussian filter attenuates these high frequency aberrations at 2 bpc, leading to a blurred representation with `good' visual quality (a score between 60-80~\cite{500}).

There is no significant difference in quality at either 6 or 8 bpc compared to the 10 bpc reference for any of the tested sequence or methods. Therefore for the given experimental setup, a bit depth of 8 or 10 bits is predicted to be perceptually redundant. While the contrast ratio used in this experiment (400:1) is fairly low compared to HDR displays (which can be upwards of 20,000:1), our results indicate that, beyond a certain point (the critical bit depth), increased bit depth representations produce no gain in visual quality. This suggests that bit depth adaptation is a viable method for reducing overhead and, when coupled with a suitable quality metric, can be used for content dependent bit depth selection within a codec. Further investigation is though required to fully understand interactions between reduced bit representations, adaptation methods and video compression processes.

%Given that the display in the experiment had a contrast ratio of 400:1 and a critical bit depth of around 6 bpc, a value of 0.16 pixel values per contrast ratio is approximated. 12 bpc would then be predicted for a HDR display with a contrast ratio of 20,000:1, and while this is based on a number of assumptions, it is consistent with current video standards~\cite{2020}.

\subsection{Viewer Preference}
\label{subsec:preference}

Overall viewer preferences at each tested bit depth are shown in Fig.~\ref{fig:VP}. While there is no clear preference at 6 and 8 bpc (above the critical bit depth), there is a slight preference for error diffusion at 4 bpc and a clear preference for the adaptive Gaussian method at 2 bpc. Error diffusion appears to be the optimal method in terms of complexity and quality, as it is the preferred method across most of the tested bit depths (apart from when very low bit depths are employed).

\subsection{Evaluation of Popular Quality Metrics}

 Any method employed in future adaptive formats would be dependent on factors including the available channel bandwidth, display parameters and content type. In this context, a quality assessment algorithm which is robust to content dependency and the bit depth adaptation method used, while providing accurate quality predictions, is required. Therefore we have tested a number of popular image quality metrics on the subjective evaluations collected to scrutinize performance across the three tested bit depth adaptation methods.

\begin{table}[t]
\footnotesize
\setlength{\tabcolsep}{-2pt}
\renewcommand\arraystretch{0}
  \centering  
\vspace{-1.5ex}
  \begin{tabular}{c}
	\includegraphics[width=0.88\columnwidth]{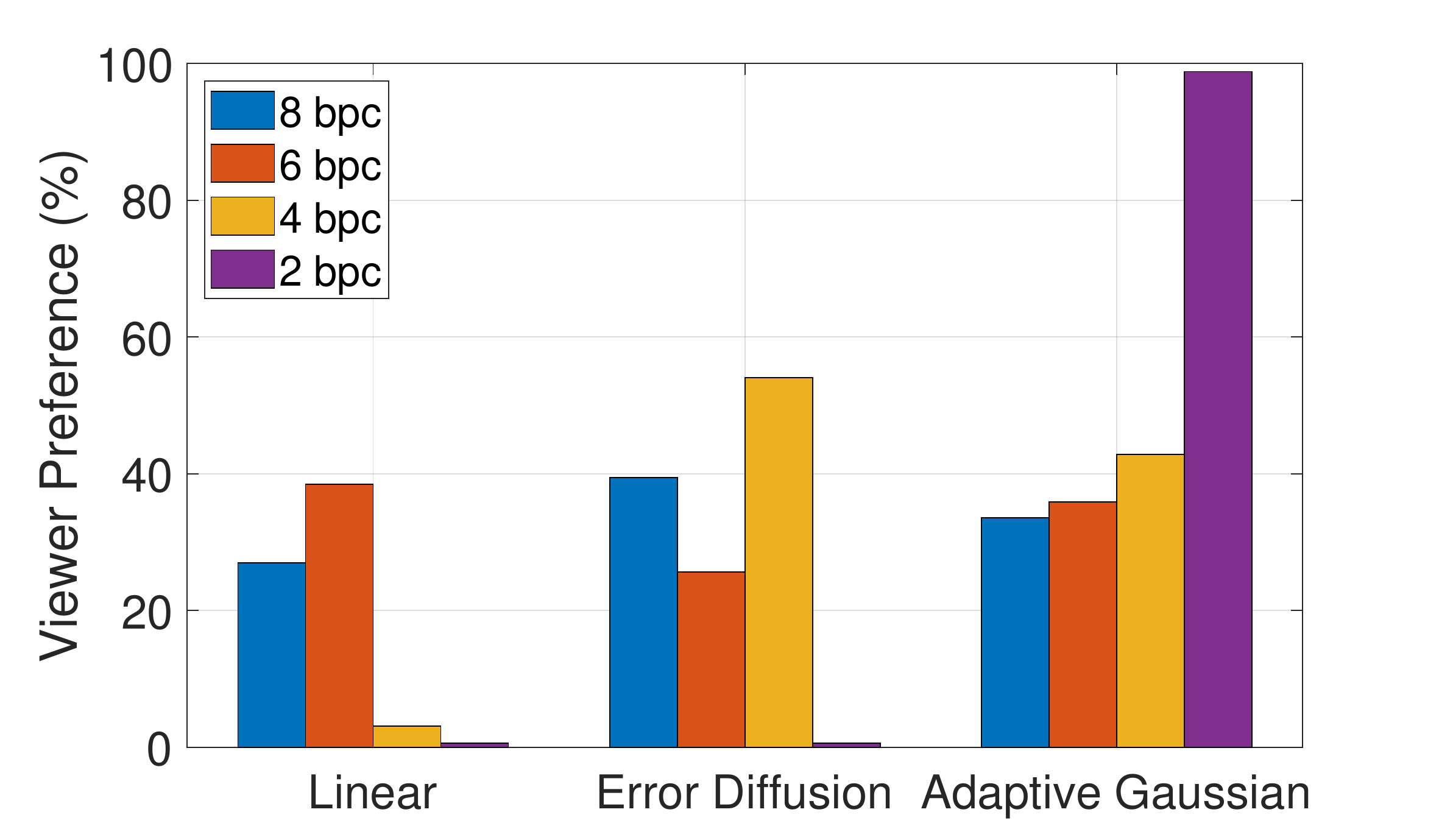}\\
\end{tabular}
\vspace{-1.5ex}
\captionof{figure}{Viewer preference for each adaptation method.}
\label{fig:VP}
\end{table}

The image quality metrics tested in this paper include: PSNR \cite{winkler2008evolution}, SSIM \cite{wang2004image}, MS-SSIM \cite{wang2003multiscale}, VSNR \cite{chandler2007vsnr}, VIF \cite{sheikh2006image}, and VMAF \cite{li2016toward}. Video quality metrics such as HDR-VQM~\cite{narwaria2015hdr} have been intentionally omitted due to complexity. 

Predictions were averaged over all frames in each sequence, and separately fitted to the differential mean opinion scores (DMOS:  MOS difference between the reference video and the low bit depth version) using a logistic function. Four  correlation statistics- Spearman Rank Correlation (SROCC), Pearson Linear Correlation (LCC), Outlier Ratio (OR) and Root Mean Squared Error (RMSE) are used to characterize the overall correlation performance of these metrics \cite{seshadrinathan2010study}.

\begin{table}[htbp]
\footnotesize
\centering
\caption{The performance of the tested quality metrics on the BVI-BD database. The best performer is highlighted in \textbf{bold}.}
\begin{tabular}{c | c | c| c| c | c| c}
\toprule
\multirow{1}{*}{} & \multicolumn{1}{c|}{PSNR} & \multicolumn{1}{c|}{SSIM}&\multicolumn{1}{c|}{MS-SSIM}&\multicolumn{1}{c|}{VSNR}&\multicolumn{1}{c|}{VMAF}&\multicolumn{1}{c}{VIF}\\
\midrule
SROCC & 0.319 & 0.703 & 0.788 & 0.765 & 0.728 & \textbf{0.830}\\
\midrule 
LCC & 0.468 & 0.782 & 0.887 & 0.826 & 0.626 & \textbf{0.910}\\
\midrule 
OR & 0.625 & 0.458 & 0.281 & 0.311 & 0.608 & \textbf{0.264}\\
\midrule 
RMSE & 28.956 & 20.703 & 14.723 & 18.134 & 24.336 & \textbf{13.047}
\\\bottomrule
\end{tabular}
\label{tab:QM}
\end{table}

% \begin{table}[htbp]
% \scriptsize
% \centering
% \caption{The performance of the tested quality metrics on the BVI-BD database based on the three bit depth adaptation filters. The best performer is highlighted in \textbf{bold}.}
% \begin{tabular}{l| r | r| r| r }
% \toprule
% \multirow{1}{*}{Metric} & \multicolumn{1}{c|}{SROCC} & \multicolumn{1}{c|}{LCC}&\multicolumn{1}{c|}{OR}&\multicolumn{1}{c}{RMSE}\\
% \midrule
% PSNR & 0.319 & 0.468 & 0.625 & 28.956\\
% \midrule 
% SSIM & 0.703 & 0.782 & 0.458 & 20.703\\
% \midrule 
% MS-SSIM & 0.788 & 0.887 & 0.281 & 14.723\\
% \midrule 
% VSNR & 0.765 & 0.826 & 0.311 & 18.134\\
% \midrule 
% VMAF & 0.728 & 0.626 & 0.608 & 24.336\\
% \midrule 
% VIF & \textbf{0.830} & \textbf{0.910} & \textbf{0.264} & \textbf{13.047}
% \\\bottomrule
% \end{tabular}
% \label{tab:QM}
% \end{table}

Table.~\ref{tab:QM} shows the performance of the tested quality metrics using BVI-BD across the three bit depth adaptation method. VIF achieves the best performance in every statistic.

%\begin{figure}[h]
%\centering
%\scriptsize
%\centering
%\begin{minipage}[b]{0.45\linewidth}
%\centering
%\centerline{\includegraphics[width=1.12\linewidth]{PSNR.pdf}}
%\end{minipage}
%\begin{minipage}[b]{0.45\linewidth}
%\centering
%\centerline{\includegraphics[width=1.12\linewidth]{SSIM.pdf}}
%\end{minipage}

% \begin{minipage}[b]{0.45\linewidth}
% \centering
% \centerline{\includegraphics[width=1.12\linewidth]{MSSSIM.pdf}}
% \end{minipage}
% \begin{minipage}[b]{0.45\linewidth}
% \centering
% \centerline{\includegraphics[width=1.12\linewidth]{VSNR.pdf}}
% \end{minipage}

% \begin{minipage}[b]{0.45\linewidth}
% \centering
% \centerline{\includegraphics[width=1.12\linewidth]{VIF.pdf}}
% \end{minipage}
% \begin{minipage}[b]{0.45\linewidth}
% \centering
% \centerline{\includegraphics[width=1.12\linewidth]{VMAF.pdf}}
% \end{minipage}

% \caption{Scatter plots of DMOS versus different quality metrics based on the three bit depth adaptation filters.}
% \label{fig:QM}
% \end{figure}

\section{Conclusion}
\label{sec:conclusion}

This paper characterizes the effect of bit depth on visual quality and investigates three methods for bit depth adaptation. Alongside linear scaling and error diffusion, we propose a novel method based on an adaptive Gaussian filter. While simple, this method demonstrates that post-filtering after bit depth adaptation can retain visual quality even after large reductions in bit depth. The subjective results show that there is a critical bit depth (around 6 bpc here), above which bit depth adaptation can employed to reduce overhead without impacting visual quality. Below this limit, 'good' visual quality can still be retained, dependent on content, by using advanced adaptation methods. Future work should focus on the development of a bespoke quality metric for videos with re-sampled bit depths and more comprehensive subjective investigation on content with both compression and bit depth adaptation artifacts. Further work is also required to fully integrate these approaches into conventional video compression architectures.

\fontsize{8.7}{9.8}\selectfont
\bibliographystyle{IEEEbib}
\bibliography{refs}
\end{document}